# Time-dependent fields and anisotropy dominated magnetic media

K. Rivkin[1], W. Saslow[1] and J. B. Ketterson[2,3,4]

1. Department of Physics and Astronomy, Texas A&M University
College Station TX, 77843

2. Department of Physics and Astronomy, Northwestern University
Evanston IL, 60201

3. Department of Electrical and Computer Engineering, Northwestern University
Evanston IL, 60201

4. Materials Research Center, Northwestern University
Evanston IL, 60201

**Abstract**

We use a single dipole approximation to analyze the behavior of anisotropy-dominated magnetic nanoparticles subjected to an external r.f. field. We identify the steady state oscillations and analyze their stability. We also analyze the case when the external r.f. field has a time-dependent frequency which insures the most effective switching of the magnetization.

PACS: 75.60.Jk, 75.40.Gb, 76.50.+g

**1. Introduction.**

The behavior of magnetic systems subjected to a strong r.f. field has been extensively studied since the middle of the last century. Traditionally such studies lead to the onset of the so-called Suhl instabilities[1], a series of non-linear phenomena corresponding to the excitation of spatially inhomogeneous spin waves. Such excitation occurs primarily via parametric resonance – excitation of a spin waves with a frequency equal to half that of a uniform mode, and via second harmonics generation – excitation of a spin wave with a frequency equal to twice that of a uniform mode. However, as was shown recently[2], in nanoscale systems Suhl instabilities can be suppressed: due to exchange effects the distance between modes becomes larger and one can have a situation where there are no modes available to couple to either parametric resonance or second harmonics generation. On the other hand there has been a revival in the past few years of interest in using the r.f. pulses to facilitate magnetization reversal[3,4,5]. Since modern hard drive writing heads are capable of producing extremely strong time-dependent magnetic fields it is clear that the possible advantages of exploiting the resonant properties of magnetic systems should be explored; in particular, the use of oscillatory, rather than flat-topped, magnetic pulses.

The case where both small r.f. and large d.c. field are used to facilitate the switching and, in addition, the system can be approximated by a single magnetic dipole, has been studied Bertotti, Serpico and Mayergoyz[6]. Here we analyze a different configuration, where there is no d.c. field present, but the switching is accomplished by a strong r.f. pulse acting on a small magnetic sample with uniaxial anisotropy.

In our previous work[7] we were able to establish, numerically, that switching is possible even when one uses a constant frequency r.f. field; however the required fields are relatively large. We were not able to provide an analytical model that would describe the physical nature of this process or to predict its dependence on the system parameters. We were also able to show that one can use so called "chirped" pulses with a time-dependent frequency – in this case one can reverse the magnetization with relatively small fields, and the fields required for the switching and the switching time only slightly depend on the value of the anisotropy constant. In the present manuscript we will analyze the stability[8] of such pulses and other important details of their application.

**2. Lyapunov stability of a single dipole in an r.f. field in the presence of anisotropy.**

We consider a magnetic sample with uniaxial anisotropy, which is small enough that it can be represented as a single dipole. We arrange the coordinate system in such a way that the z-axis is oriented parallel to the easy axis. The anisotropy field can then be expressed as:

$$\mathbf{h}_{ani} = \hat{z} K M_z \qquad (2.1)$$

where K is a dimensionless coefficient, $M_z$ is the z component of the magnetization. We assume the external time dependent magnetic field has the circularly-polarizecd form:

$$\mathbf{h}^{rf} = \hat{x} H_1 \cos(\varphi_h(t)) + \hat{y} H_1 \sin(\varphi_h(t)) \qquad (2.2)$$

where $H_1$ is the amplitude and $\varphi_h(t)$ is a time dependent phase. Precession of a single dipole in the presence of anisotropy, damping (governed by the damping coefficient $\beta$) and external r.f. field is then given by the Landau-Lifshitz[9] equation:

$$\frac{d\mathbf{m}}{dt'} = -\gamma \mathbf{m} \times \mathbf{h}^{total} - \frac{\beta\gamma}{M_s} \mathbf{m} \times \left(\mathbf{m} \times \mathbf{h}^{total}\right)$$
$$\mathbf{h}^{total} = \mathbf{h}_{ani} + \mathbf{h}^{rf}.$$
(2.3)

We apply the following transformation of the time and field variables:

$$t = t'\gamma K M_s$$
$$H = \frac{H_1}{KM_s}$$
(2.4)

Then, in spherical coordinates $(\theta, \varphi, r)$, one can rewrite equation (2.3) as:

$$\dot{\theta} = H \sin \Delta\varphi - \beta \sin\theta \cos\theta$$
$$\Delta\varphi = \varphi_h - \varphi$$
$$\Delta\dot{\varphi} = \dot{\varphi}_h - \cos\theta + H \cos\Delta\varphi \frac{\cos\theta}{\sin\theta}.$$
(2.5)

Because of the symmetry, the system behavior depends only on $\Delta\varphi$ - the difference between the phase of the r.f. field, $\varphi_h(t)$ and the azimuthal angle, $\varphi$. Initially the magnetic moment is aligned parallel to z axis, $\theta = 0$.

We now examine the basic linear stability of the system using Lyapunov's analysis.[7] We suppose that for some trajectory given by Eqs. (2.5), the motion is perturbed at a point $\theta, \Delta\varphi$ (e.g., by a finite temperature). Expanding Eqs. (2.5) in Taylor's series with respect to a small perturbation $\begin{pmatrix} \delta\theta \\ \delta\Delta\varphi \end{pmatrix}$ and retaining only the linear term one obtains:

$$\begin{pmatrix} \dfrac{d\delta\theta}{dt} \\ \dfrac{d\delta\Delta\varphi}{dt} \end{pmatrix} = \begin{pmatrix} \dfrac{d(H\sin\Delta\varphi - \beta\sin\theta\cos\theta)}{d\theta} & \dfrac{d(H\sin\Delta\varphi - \beta\sin\theta\cos\theta)}{d\Delta\varphi} \\ \dfrac{d\left(\dot{\varphi}_h - \cos\theta + H\cos\Delta\varphi\dfrac{\cos\theta}{\sin\theta}\right)}{d\theta} & \dfrac{d\left(\dot{\varphi}_h - \cos\theta + H\cos\Delta\varphi\dfrac{\cos\theta}{\sin\theta}\right)}{d\Delta\varphi} \end{pmatrix}\Bigg|_{\theta,\Delta\varphi} \begin{pmatrix} \delta\theta \\ \Delta\varphi \end{pmatrix}.$$

(2.6)

Writing

$$\begin{pmatrix} \delta\theta(t) \\ \delta\Delta\varphi(t) \end{pmatrix} = \begin{pmatrix} \delta\theta \\ \delta\Delta\varphi \end{pmatrix}\Bigg|_{t=0} e^{\lambda t},$$

(2.7a)

Eq. (2.6) takes the form of an eigenvalue equation:

$$\lambda \begin{pmatrix} \delta\theta(t) \\ \delta\Delta\varphi(t) \end{pmatrix} = \begin{pmatrix} -\beta\cos 2\theta & H\cos(\Delta\varphi) \\ \sin\theta - H\cos\Delta\varphi\left(\dfrac{1}{\sin^2\theta}\right) & -H\dfrac{\cos\theta}{\sin\theta}\sin(\Delta\varphi) \end{pmatrix}\Bigg|_{\theta,\Delta\varphi} \begin{pmatrix} \delta\theta \\ \delta\Delta\varphi \end{pmatrix}.$$

(2.7b)

Solving Eq. (2.7b) for the eigenvalues $\lambda_{1,2}$ we obtain:

$$\begin{aligned} a &= \frac{1}{2}\left(H\frac{\cos\theta\sin(\Delta\varphi)}{\sin\theta} + \beta\cos 2\theta\right) \\ b &= \beta H\frac{\cos 2\theta\cos\theta\sin(\Delta\varphi)}{\sin\theta} - H\sin\theta\cos(\Delta\varphi) + \left(\frac{H\cos\Delta\varphi}{\sin\theta}\right)^2 \\ \lambda_1 &= -a + \sqrt{a^2 - b} \\ \lambda_2 &= -a - \sqrt{a^2 - b} \end{aligned}$$

(2.8)

Stability of the system is determined by the sign of the real part of the largest eigenvalue, $\lambda_1$. If it is positive, any small perturbation exponentially diverges; the trajectory is unstable, and the system has exhibits chaotic behavior, i.e. the trajectory becomes sensitive to small perturbation of initial conditions, and, as it will be shown later, formation of a chaotic attractor is possible for a specific form of the applied r.f. field. If $\lambda_1$ is negative, any trajectory obtained by a small deviation from the initial trajectory decays back to the original trajectory.

We now ask what regions of the $\theta, \Delta\varphi$ phase space lead to stable behavior. In Figure 1 we give a three-dimensional plot of $\text{Re}(\lambda_1)$ as a function of $\theta$ and $\Delta\varphi$ for the case $\beta = 0.01,\ H = 0.2$. Red corresponds to very large positive values, blue to very small negative values. Lines connect the points with the same value of $\text{Re}(\lambda_1)$. In Figure 2 we specifically show the areas where $\text{Re}(\lambda_1) < 0$ in black and $\text{Re}(\lambda_1) > 0$ in white. A flattening of some of the boundaries between these regions arises from numerical round-off errors. It is clear from the symmetry that if one knows the stability for $0 \leq \theta \leq \pi/2$ and $\pi/2 \leq \Delta\varphi \leq \pi$, one can construct $\lambda_1$ for the entire space, due to the intrinsic symmetry properties of the system. It is possible to obtain analytical expressions for the boundary between stable and unstable regimes (Fig.2) by solving the Eq. (2.8) for $\text{Re}(\lambda_1) = 0$. It can be shown that for relatively small damping there are two distinctive solutions.

In the first case $\dfrac{\pi}{2} \leq \Delta\varphi < \pi$ and the system becomes unstable when a is negative:

$$a \leq 0 \Rightarrow \theta \geq \frac{\pi}{2} \tag{2.9a}$$

However this is a very approximate solution and, as will be shown below, a more careful analysis is needed when $\Delta\varphi$ is close to $\pi$.

In the second case, $0 \leq \Delta\varphi \leq \frac{\pi}{2}$, where the system becomes unstable if a is positive and b is negative, we have

$$b \leq 0 \Rightarrow \sin^3\theta \geq H\cos(\Delta\varphi) \tag{2.9b}$$

## 3. Switching with a constant frequency r.f. field.

We now analyze the case when $\dot{\varphi}_h = \omega = \text{constant}$, i.e. the case where the frequency of the applied r.f. field is time-independent. The first question we ask is whether it is possible for such a system to have a steady state behavior, i.e., can we have

$$\begin{aligned}0 &= H\sin\Delta\varphi - \beta\sin\theta\cos\theta \\ 0 &= \omega - \cos\theta + H\cos\Delta\varphi \frac{\cos\theta}{\sin\theta}\end{aligned} \tag{3.1}$$

Numerically solving Eq. (2.10) we see that there is always at least one stable equilibrium configuration present in the system. For $\omega > 0$ we always have equilibrium with $\theta = \pi$, i.e. the magnetic moment is parallel to the easy axis, but anti-parallel to the initial configuration $\theta = 0$. This is due to the fact that the resonant frequency of such a configuration is $-1$ in the units given by Eq. (2.4), while the resonant frequency of the $\theta = 0$ configuration is $+1$, i.e. the system moves as far away from the resonance as possible. This and other equilibrium configurations correspond to $\Delta\varphi$ being close to either 0 or $\pi$, where the external r.f. field produces only a very small torque. In Figure (3) we show different equilibrium values of $\theta$ as a function of $\omega$ for a fixed H = 0.2. The black curve corresponds to the stable equilibrium with the magnetic moment anti-parallel to the initial configuration. The red curve ($\Delta\varphi \approx \pi$) shows the equilibrium occurring for sufficiently small values of H, and for $\omega$ close to 1, corresponds to linear oscillations around the initial configuration. It becomes unstable for sufficiently high values of H and small values of $\omega$. Green and blue curves show symmetric solutions, the first of which is unstable ($\Delta\varphi \approx \pi$) and another is stable ($\Delta\varphi \approx 0$). The reason underlying this symmetry is that $\sin\theta\cos\theta$ is the same for $\theta = \pi/4 + \delta\theta$ and $\theta = \pi/4 - \delta\theta$, which creates two solutions in the vicinity of $\theta = \pi/4$.

By solving Eqs. (2.8) and (3.1) given $\lambda_1 = 0$ we can find the instability condition for the equilibrium given by a blue curve ($\Delta\varphi \approx 0$)

$$\begin{aligned}&\Delta\varphi \approx 0 \\ &\sin^3\theta \geq H \\ &\omega - \cos\theta + \sin^2\theta\cos\theta \approx 0 \\ &\omega - \cos^3\theta \approx 0\end{aligned} \quad (3.2)$$

which leads to:

$$H > \left(1 - \omega^{\frac{2}{3}}\right)^{\frac{3}{2}}. \quad (3.3)$$

The instability condition for the equilibrium given by a red curve ($\Delta\varphi \approx \pi$) can be obtained by solving:

$$\begin{aligned}&\Delta\varphi \approx \pi \\ &a = \frac{1}{2}\left(H\frac{\cos\theta\sin(\Delta\varphi)}{\sin\theta} + \beta\cos 2\theta\right) = 0 \\ &\omega - \cos\theta + H\cos\Delta\varphi\frac{\cos\theta}{\sin\theta} = 0\end{aligned} \quad (3.4)$$

which gives the following critical value of H, above which the solution becomes unstable:

$$H = \sqrt{2}\omega - \sqrt{\frac{2}{3}}.\tag{3.5}$$

In Figure (4) we show how the critical fields given by Eqs. (3.4) and (3.5) depend on H. Obviously, if H is large enough, both of these equilibriums are unstable. The smallest such H corresponds to the frequency when both these equilibriums become unstable at the same H:

$$\sqrt{2}\omega - \frac{\sqrt{2}}{\sqrt{3}} = \left(1 - \omega^{\frac{2}{3}}\right)^{\frac{3}{2}},\tag{3.6}$$

$$\omega \approx \frac{2}{3}, \quad H \approx 0.126.$$

We now ask what happens when H is greater or smaller the value for which these equilibriums become unstable. Using a 4$^{th}$ order Runge-Kutta algorithm we can solve Eq. (2.5) numerically. In Figure 5 we show the behavior when H is smaller than this critical value – the system converges to one of the equilibriums with $\theta < \frac{\pi}{2}$. If H is greater than the critical value, as in Figure 6, the system converges to the equilibrium located around $\theta = \pi$. Convergence is extremely slow, since the system spends a lot of time in the areas where the derivatives in Eq. (2.5) are very close to 0 and the system is highly unstable ($\text{Re}(\lambda) > 0$).

As one can see, this scenario is very different from the case where one applies a small r.f. field perpendicular to a large d.c. field. Instead of looking for the conditions corresponding to a maximum absorption per unit time, when only anisotropy and an r.f. field are present the only way to effectively switch the magnetization is to select values of $\omega$ and H such that there are no stable equilibriums in the vicinity of the initial direction of the magnetization.

## 4. "Chirped" pulses.

We now consider the case where, instead of $\dot{\varphi}_h = \omega = \text{constant}$, one tries to maintain a constant $|\Delta\varphi| \approx \frac{\pi}{2}$. As seen from Eq. (2.5) this corresponds to the most rapid change of $\theta$ and therefore the fastest switching of the magnetization. One of the simpler pulses that accomplishes this has been published earlier and corresponds to[7]

$$\dot{\varphi}_h = \cos Ht$$

$$\theta \approx \pm Ht ; \qquad (4.1)$$

clearly the sign of $\Delta\varphi$, and therefore the choice between $\theta \approx +Ht$ and $\theta \approx -Ht$, is irrelevant – both correspond to the same physical motion. Here we analyze the stability of the solution given by (4.1). In Figure 7 we show the results of a numerical analysis of a system subject to a "chirped" pulse. As can be seen, the process is stable for $|\theta| < \pi/2$. Any perturbation applied to the system is unlikely to move the system away from the stable path. As a result even relatively strong perturbations, like white noise with an amplitude of 0.3 radians cannot negate the switching – they cannot move the system to another stable path and on the average the system remains in the vicinity of a stable solution given by Eq. (4.1). The same can be said about the initial conditions – if for example we start with $\Delta\varphi(t=0)$ far from $\frac{\pi}{2}$, in an extremely short time the system assumes the dynamics such that $\Delta\varphi = \pm\frac{\pi}{2}$.

For $|\theta| > \frac{\pi}{2}$ the situation is drastically different. Suppose initially we followed the path where $\Delta\varphi = \frac{\pi}{2}$ (as was mentioned before, precisely the same physical situation arises if $\Delta\varphi = -\frac{\pi}{2}; 0 > \theta$). If $\theta > \frac{\pi}{2}$ such a solution becomes intrinsically unstable, while another solution with $\Delta\varphi = -\frac{\pi}{2}$ becomes increasingly stable. If the system had proceeded along the $\Delta\varphi = \frac{\pi}{2}$ solution, $\theta$ would linearly increase with time until it became greater than $\pi$, which would mean that the system completely switched and was now switching back towards the original state $\theta = 0$. If there would be no stability considerations, such switching would continue back and forth for forever. However, as we have said, in reality for $\theta > \frac{\pi}{2}$ such a solution is unstable, and the system moves from an unstable to a stable solution, which is $\Delta\varphi = -\frac{\pi}{2}; \theta > \frac{\pi}{2}$. Such a move means that the system does not switch completely, i.e., it does not reach $\theta = \pi$, but instead reaches some value of $\theta$ greater than $\frac{\pi}{2}$ and than $\theta$ start to decrease towards 0. In Figures 7 and 8 we show the first few cycles – the system starts with $\Delta\varphi = \frac{\pi}{2}; \theta = 0$, reaches $\theta > \frac{\pi}{2}$, moves to a more stable orbit where $\Delta\varphi = -\frac{\pi}{2}$, $\theta$ decreases almost to 0, the system again acquires $\Delta\varphi = +\frac{\pi}{2}$, and so on. During the first few periods determined by the value of H, the maximum value of $\theta$ reached is very close to $\pi$; i.e., even though the motion is formally unstable in the angular range $\pi/2 < \theta < \pi$ when $\Delta\varphi = +\dfrac{\pi}{2}$, such chirped pi pulses can reverse the magnetization (in any case, on removal of the pulse when $\theta > \pi/2$ the sample will precess down to $\theta = \pi$).

Interesting behavior can be observed after thousands of periods. Keeping in mind that the applied r.f. field is periodic, we see that the system displays a quasi periodic behavior (Figure 9); it approximately reproduces itself over and over again, even in the presence of small perturbations. However, since the motion includes large segments where it is unstable (as can be seen from Lyapunov eigenvalues shown in Figure 10) we conclude it is chaotic but, due to its quasi periodic behavior, one possessing a chaotic attractor. We can construct a Poincare map as shown in figure 11 by recording values of $\Delta\varphi$ and $\theta$ *separated by the period of the applied r.f. field* $-2\pi/H$; note that the true motion is much more complicated than the one shown in Figure 9 – the system does not spend time in just a portion of phase space corresponding to $\theta > \pi/2$ but slowly precesses on a complex and partially unstable trajectory.

**Conclusions**

In the present work we were able to establish the following:

a. Unlike the case when both d.c. and r.f. fields are used, the switching with r.f. fields alone is bound to involve chaotic behavior.

b. In the presence of an external r.f. field with a constant frequency, the anisotropy-dominated system slowly reaches, due to the damping, a steady-state precession state characterized by a constant angle between the magnetization and the easy axis with a constant phase difference between magnetization and the external r.f. field. We derived expressions giving the stability limits of such oscillations.

c. For sufficiently high values of the external r.f. field, the only steady state is a precession nearly parallel to the easy axis, but opposite to the initial configuration. The minimal value of external r.f. field for which this is possible is $H \approx 0.126 H_{ani}$ when the frequency of the r.f. field is approximately two thirds that of the resonant frequency of small angle oscillations. Larger r.f. fields will produce magnetization reversal; however such transition will be highly chaotic and slow.

d. One can design pulses with a time-dependent frequency that very effectively switch the magnetization of the sample. At least the first reversal is resistant to large perturbations and moves the system through the unstable areas of the phase space. After many reversals the magnetization starts to precess on a complex trajectory that forms a chaotic attractor.

The program used for these simulations is available for the public use at www.rkmag.com. This work was supported by the National Science Foundation under grant ESC-02-24210.

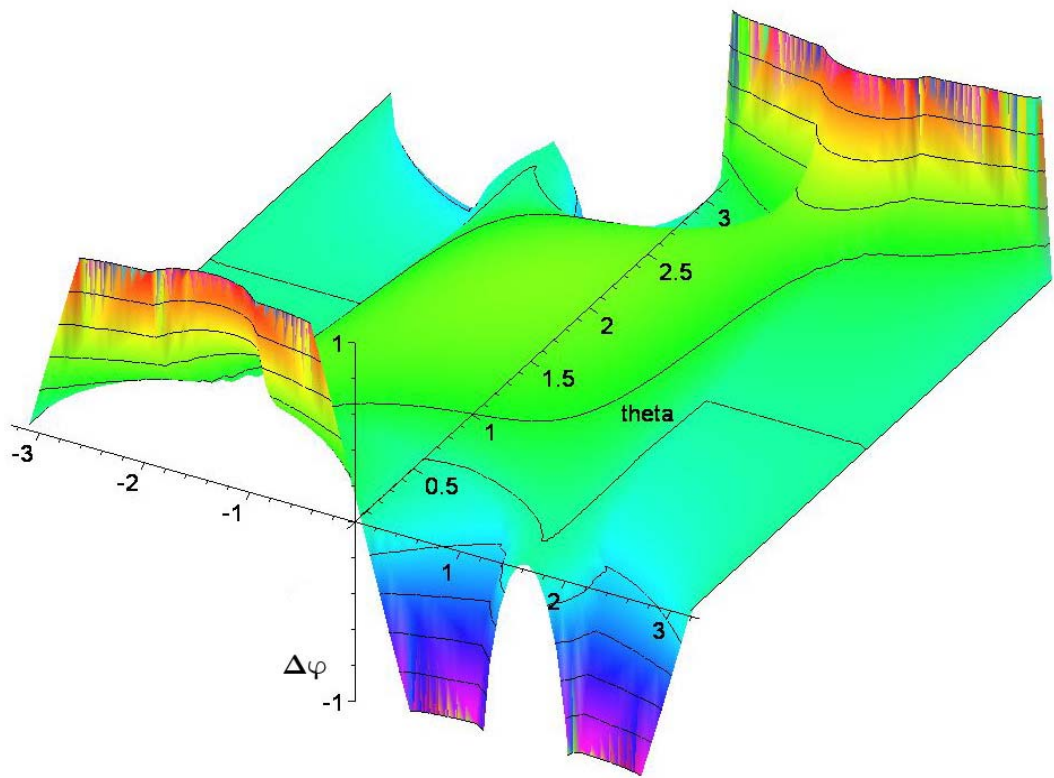

**Figure 1. Maximumal real value of Lyapunov eigenvalue as a function of** $\theta$ **and** $\Delta\varphi$.

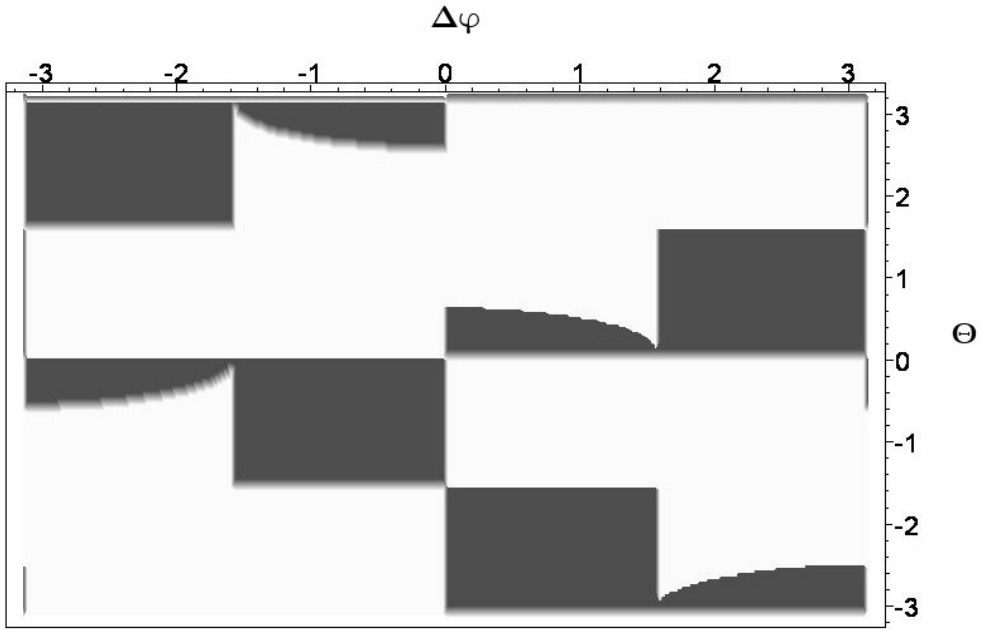

**Figure 2 . Areas in which the system is stable (black) or unstable (white).**

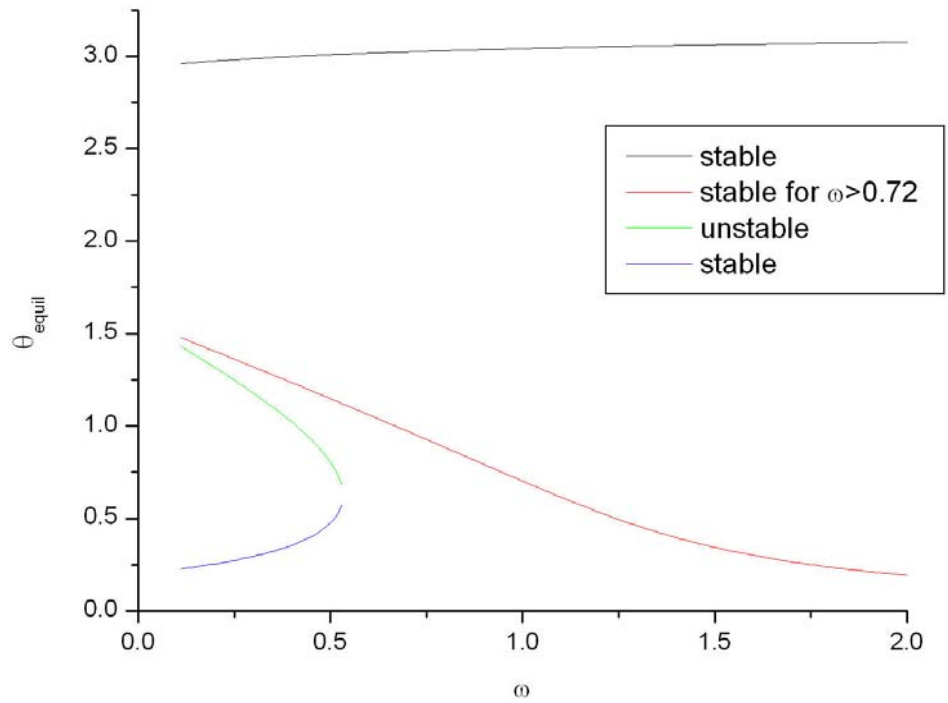

**Figure 3. Equilibrium values of θ as a function of ω for a fixed H = 0.2.**

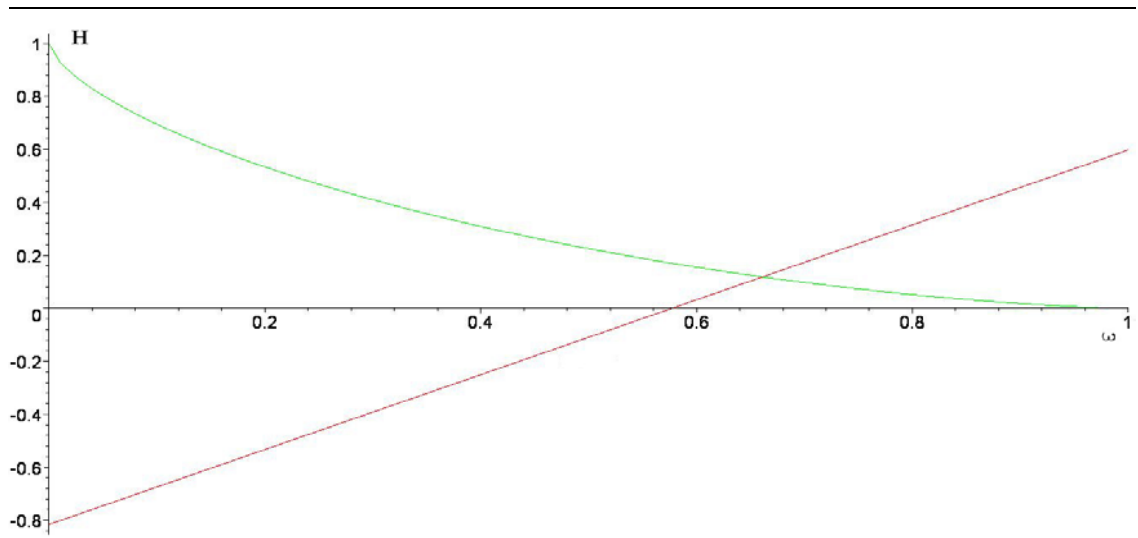

**Figure 4. Critical values of H as a function of $\omega$.**

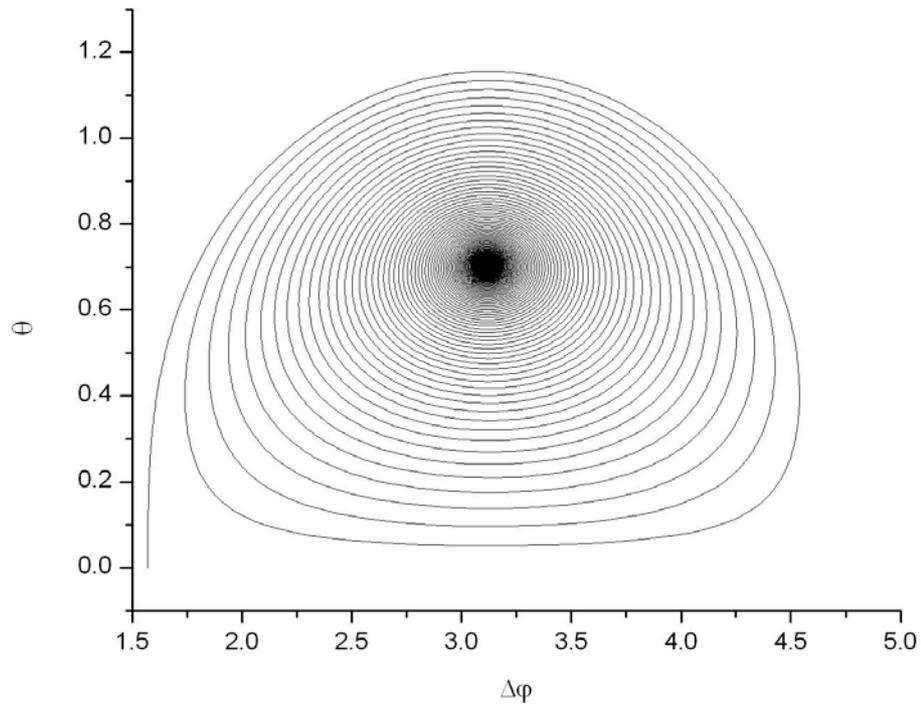

**Figure 5.** θ as a function of $\Delta\varphi$, $\omega = 1$, $H = 0.2$.

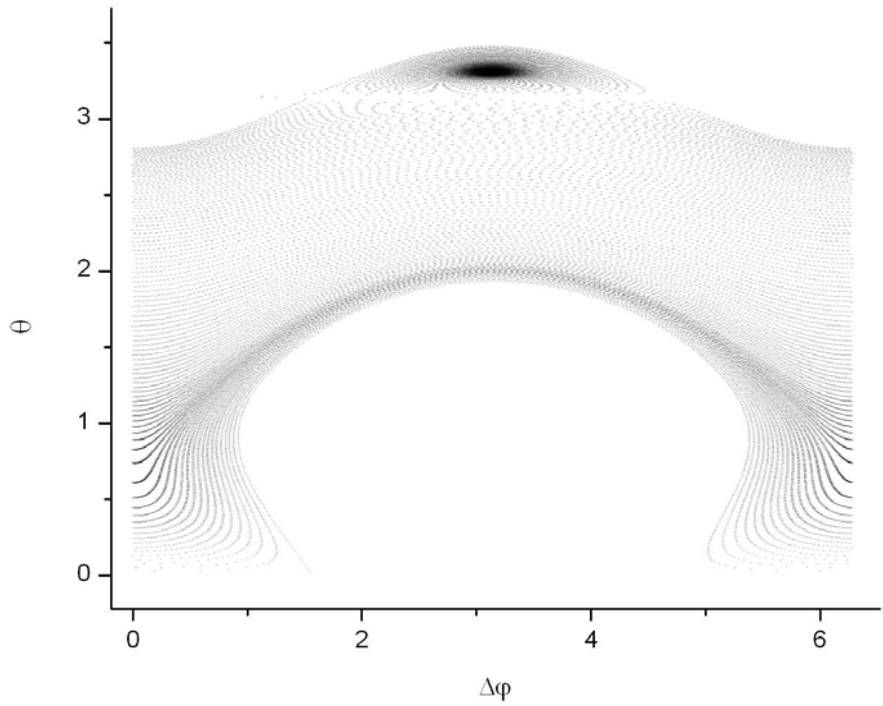

**Figure 6.** $\theta$ as a function of $\Delta\varphi$, $\omega = 0.5$, $H = 0.25$.

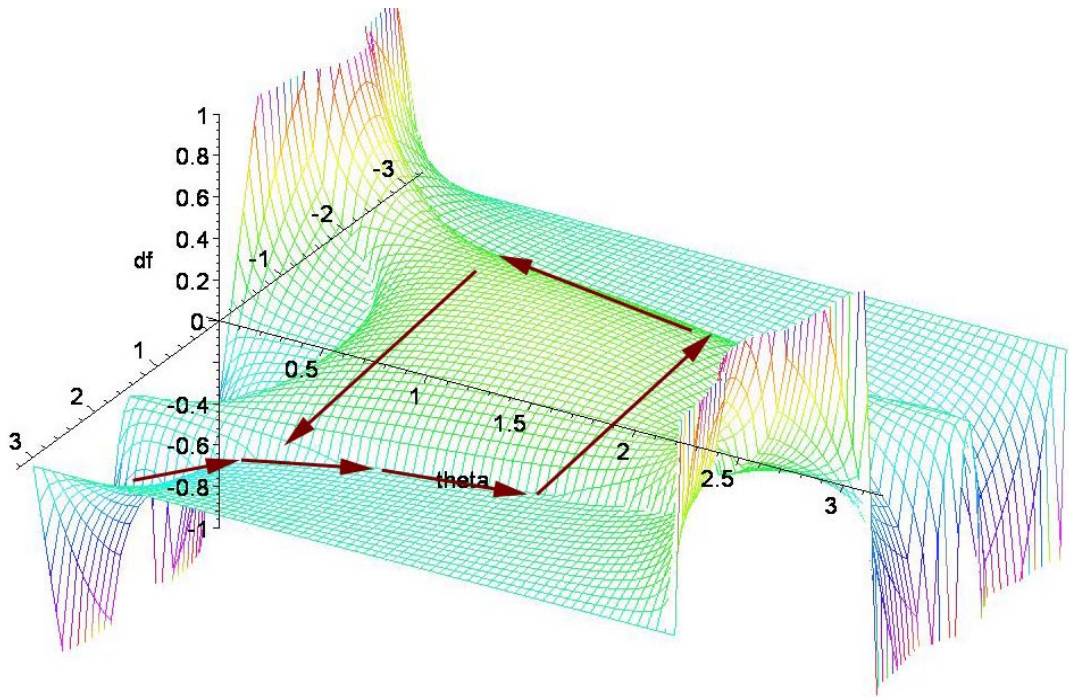

**Figure 7. Stability of magnetization dynamics in the presence of a "chirped" pulse.**

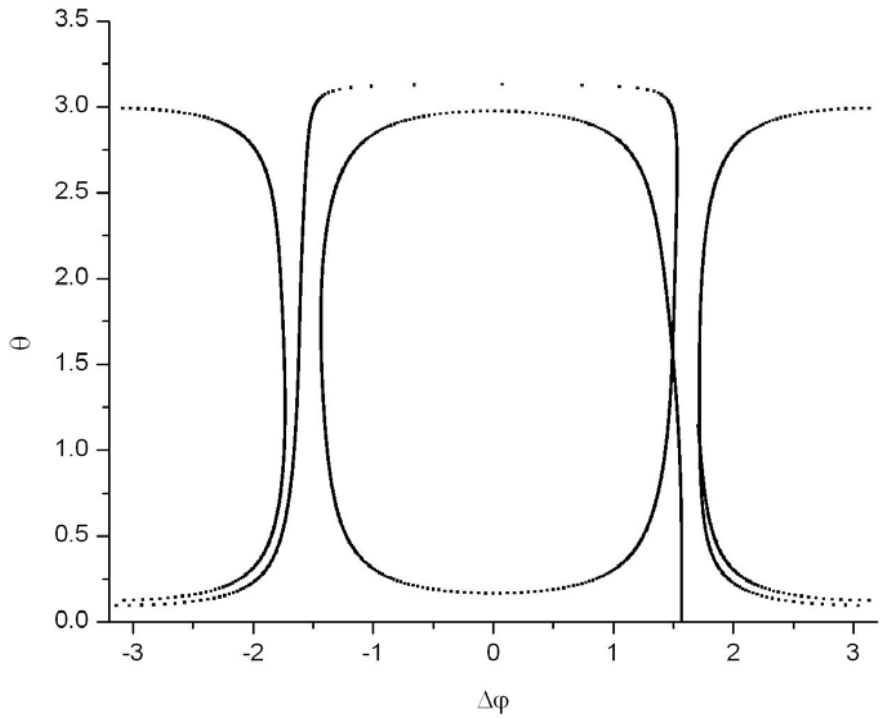

**Figure 8.** θ as a function of $\Delta\varphi$, "chirped pulse", small t, H=0.25, $\beta = 0.01$ .

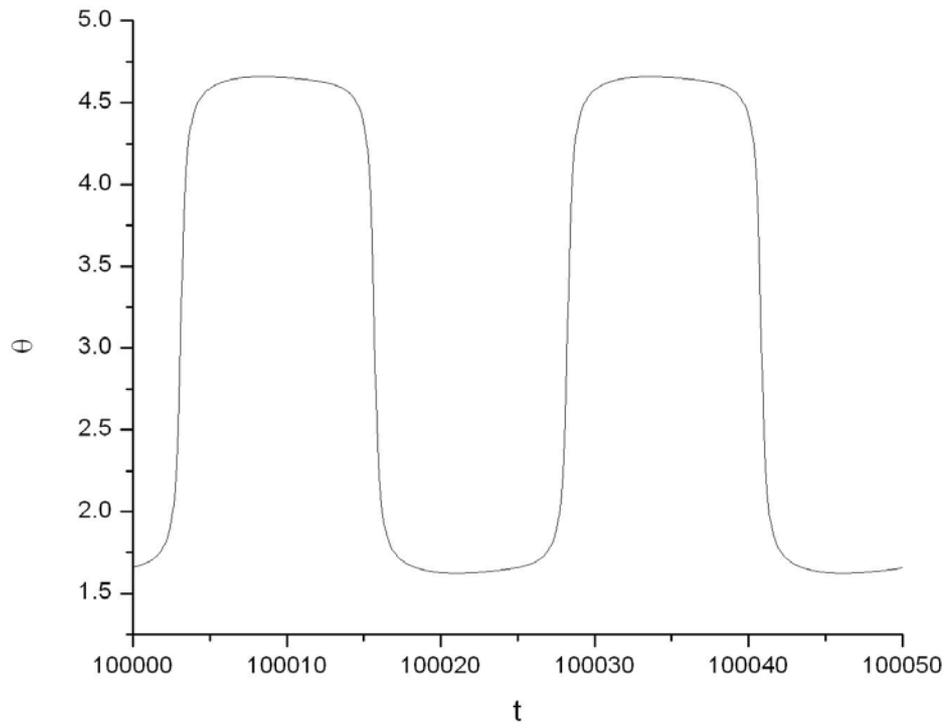

**Figure 9.** θ as a function of time, "chirped pulse", large t, H = 0.25, $\beta = 0.01$.

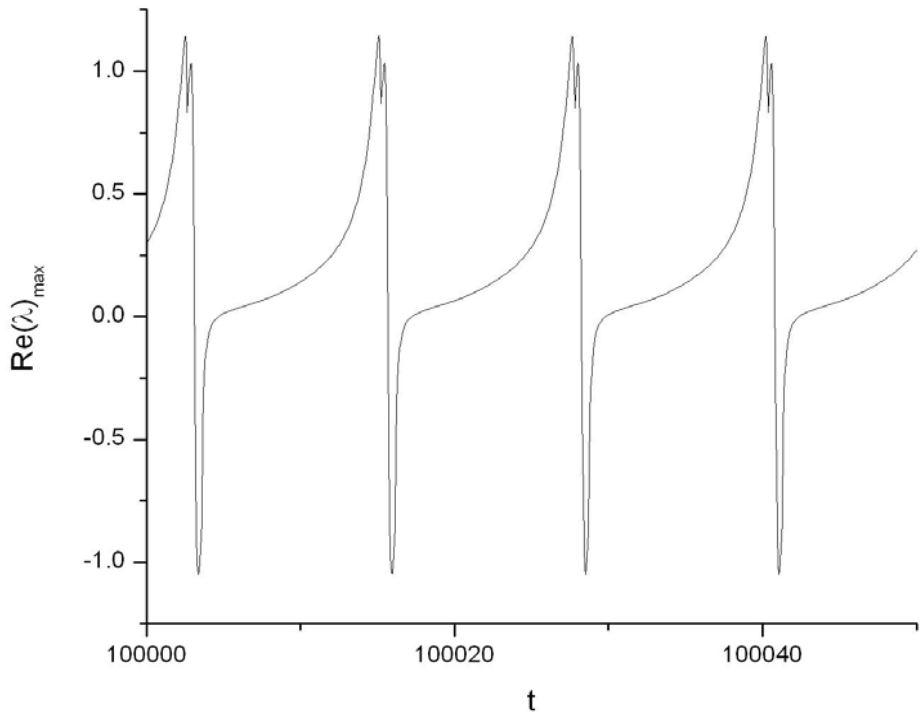

**Figure 10. Largest eigenvalue as a function of time for a "chirped pulse" at large t, for H = 0.25, β = 0.01 .**

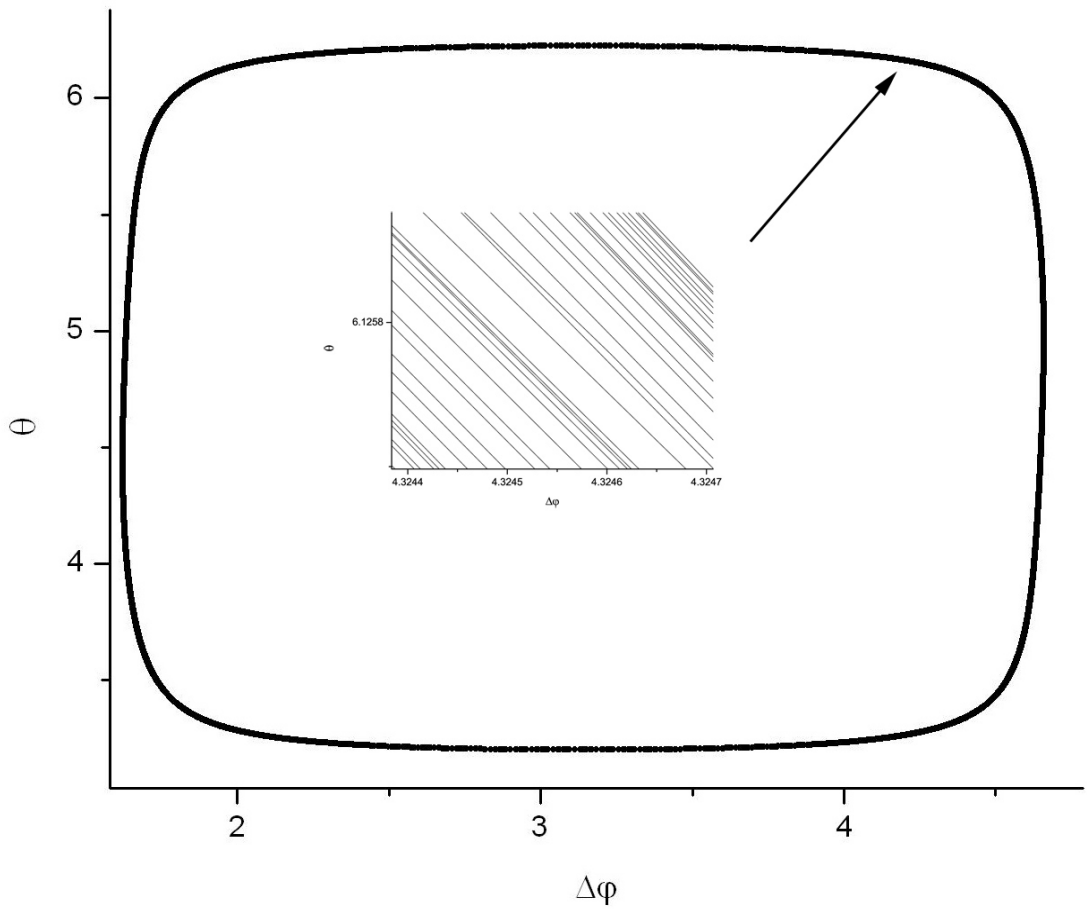

**Figure 11. Poincare map, "chirped pulse", H = 0.25, $\beta = 0.01$. The inset shows a highly magnified portion of the nominal trajectory.**